\documentclass[5p]{elsarticle}

\usepackage{lineno}
\usepackage{graphicx,graphics}   %       for graphics
\usepackage{latexsym}   %       for special symbols
\usepackage{mathrsfs}
\usepackage{fancyhdr}
\usepackage{amssymb}
\usepackage{amsfonts}
\usepackage{setspace}
\usepackage{float}
\usepackage{amsmath}
\usepackage{bm}

\modulolinenumbers[5]

\usepackage{natbib,twoopt}
\usepackage{hyperref}
\hypersetup{
    colorlinks=true,
    linkcolor=red,
    citecolor=blue,
    filecolor=magenta,      
    urlcolor=cyan,
}  %% to set colour pattern for hyperref
\begin{document}
%==============================================================================

%==============================================================================
\title{Effects of a non-zero strangeness-chemical potential in strong interaction models}
%WT = Working Title

\author{Ayon Mukherjee$^{1,2}$}

\author{Abhijit Bhattacharyya$^3$ and Stefan Schramm$^{1,2}$}

%\author{Stefan Schramm$^{1,2}$}

\address{$^{1}$Frankfurt Institute for Advanced Studies, Ruth-Moufang-Stra{\ss}e 1, D-60438 Frankfurt am Main, Germany}
\address{$^{2}$Institut f\"ur Theoretische Physik, Goethe Universit\"at Frankfurt, Max-von-Laue-Stra{\ss}e 1, D-60438 Frankfurt am Main, Germany}
\address{$^{3}$Department of Physics, University of Calcutta, 92, Acharya Prafulla Chandra Road, Kolkata 700009, India}

%\date{February 28, 2014}

\begin{abstract}
%\textcolor{blue}{$\rightarrow$ include abstract here}
The effect of a non-zero strangeness chemical potential on the strong interaction phase diagram has been
studied within the framework of the SU(3) quark-hadron chiral parity-doublet model. Both, the nuclear liquid-gas and the chiral/deconfinement phase transitions are modified. The first-order line in the chiral phase transition is observed to vanish completely, with the entire phase boundary becoming a crossover. These changes in the nature of the phase transitions are expected to modify various susceptibilities, the effects of which might be detectable in particle-number distributions resulting from moderate-temperature and high-density heavy-ion collision experiments.
\end{abstract}


%\pacs{}

\maketitle
%\section{Introduction}
%\label{sec:intro}

One of the primary foci of the ongoing, and upcoming, ultra-relativistic heavy-ion collision (HIC) experiments at RHIC in Brookhaven, at the LHC at CERN or at the future facilities, like FAIR at GSI and NICA in Dubna, is to probe the nature of strongly interacting matter under extreme conditions of temperatures and densities. However, even for these conditions strong interactions are not in the perturbative regime and are therefore extremely difficult to solve directly from first principles.

While Lattice QCD (LQCD) provides the most direct approach for studying high-temperature systems~\cite{Soltz:2015ula,Borsanyi:2013bia}, it is plagued by the familiar fermion sign-problem~\cite{Aarts:2014nxa,Aarts:2016bdr,Aarts:2017vrv,Alexandru:2015sua,Alexandru:2017dcw,Anagnostopoulos:2011cn,Anagnostopoulos:2017gos} at non-vanishing baryo-chemical potentials. Effective Lagrangian models~\cite{doi:10.1142/9789812795533_0008,Ratti:2005jh,Meisinger:1995ih,Casalbuoni1984,Csaki2002,Dhar1984,Hatsuda1994,Liu2012,Manohar1984,Rosenzweig1993,Schechter1980,Sil2004,Veneziano1982,Abu-Shady2017a,Alford2018,Banerjee2018,Atmaja2018,Mishustin:1993ub,Heide:1993yz,Papazoglou:1997uw,Papazoglou:1998vr,Tsubakihara:2009zb}, on the other hand, provide a much more tractable alternative to the study of non-perturbative, strongly-interacting matter. Following this approach, in Ref.~\cite{Mukherjee:2016nhb}, within an extended hadron-quark parity-doublet model, the QCD phase diagram and thermal fluctuations in an HIC have been studied, using the cumulants of conserved charges for a range of temperatures ($T$) and baryo-chemical potentials ($\mu_B$), at zero strangeness- ($\mu_S$) and isospin-chemical potentials ($\mu_I$).

However, the effects of a non-zero $\mu_S$ on the QCD phase-diagram and the fluctuations for low densities have been investigated in recent LQCD calculations and Hadron Resonance Gas (HRG) model (comparative) studies~\cite{Bazavov:2012jq,Borsanyi:2011sw,toublan2005qcd,Bhattacharyya:2013oya,Bellwied:2015lba,bhattacharyya2010investigation,bhattacharyya2010susceptibilities,bhattacharyya2011correlation,bhattacharyya2013thermodynamic,bhattacharyya2015fluctuation,bhattacharyya2012study,bhattacharyya2015thermodynamics}. On the other hand, the influence of a non-zero $\mu_I$ on the chiral phase transition can, in principle, be experimentally tested to some degree by varying projectile and target nuclei. It has been studied theoretically, using both effective-model and LQCD approaches~\cite{Klein:2003fy,Toublan:2003tt,nishida2004nishida,Barducci:2003un,Toublan:2004bj,Alford:1998sd}. 

From fitting observed particle ratios, $\mu_S$ has been deduced to have a value of $\sim 25 \% - 30 \%$ of $\mu_B$, while $\mu_I$ remains small, at around $2 \% - 5 \%$ of $\mu_B$~\cite{doi:10.1142/9789812795533_0008,PhysRevD.77.065016,
Becattini:1997ii,BraunMunzinger:1999qy}. These values illustrate that the strangeness- and isospin-chemical potentials, though small, are not entirely negligible. It is therefore worthwhile to study the QCD phase-diagram with non-zero isospin-, strangeness- and baryo-chemical potentials. This includes potential fluctuations in the fireball creating areas with positive and negative net-strangeness and net-isospin, respectively.

Motivated by these considerations, in this paper the authors focus on the strangeness aspect of a system at high-to-moderate temperatures and high densities. The model being used to study the effects of a non-zero $\mu_S$ (with $\mu_I$ = 0) is the Quark-Hadron Chiral Parity Doublet Model (Q$\chi$P), which is a low-energy, effective SU(3) chiral model, which has been previously used in Refs.~\cite{Mukherjee:2016nhb,Mukherjee:2017jzi}. After a brief description of the model and the slight modifications done to its parametrization, the results are presented and discussed, and 
conclusions are given. For a more detailed explanation of the model itself, see Refs.~\cite{Steinheimer:2010sp,Steinheimer:2010ib,
Steinheimer:2016cir,Steinheimer:2011ea}.

In parity-doublet formulations the Lagrangian can contain an explicit mass term for the baryons that does not
break chiral symmetry \cite{PhysRevD.39.2805}. The signature for chiral symmetry restoration is the degeneracy of the usual baryons and their respective negative-parity partner states, which are grouped in doublets $N = (N^+,N^-)$ as discussed in Refs. \cite{PhysRevD.39.2805,Hatsuda:1988mv}. Taking into account the scalar and vector condensates in mean-field approximation, the resulting Lagrangian ${\cal L}_{B}$ reads as \cite{Steinheimer:2011ea}:

\begin{eqnarray}
{\cal L}_{B} &=& \sum_i (\bar{B_{i}} i {\partial\!\!\!/} B_{i})
+ \sum_i  \left(\bar{B_{i}} m^*_{i} B_{i} \right) \nonumber \\ &+&
\sum_i  \left(\bar{B_{i}} \gamma_\mu (g_{\omega {i}} \omega^\mu +
g_{\rho {i}} \rho^\mu + g_{\phi {i}} \phi^\mu) B_{i} \right) ~,
\label{lagrangian2}
\end{eqnarray}

summing over the states of the baryon octet. The effective masses of the baryons (assuming the matter to be isospin-symmetric) are:

\begin{eqnarray}
m^*_{{i}\pm} = \sqrt{ \left[ (g^{(1)}_{\sigma {i}} \sigma + g^{(1)}_{\zeta {i}}  \zeta )^2 + (m_0+n_{s} m_{s})^2 \right]}\nonumber \\
\pm g^{(2)}_{\sigma {i}} \sigma \pm g^{(2)}_{\zeta {i}} \zeta ~,
\label{effmass}
\end{eqnarray}

with $g^{(j)}_{i}$'s as the coupling constants of the baryons with the scalar fields $\sigma$ $\left(\left<\overline{\psi} \psi \right> \right)$ and $\zeta$ $\left(\left<\overline{s} s \right> \right)$. In addition, there is an SU(3) symmetry-breaking mass term proportional to the strangeness, $n_{s}$, of the respective baryon. Note that in the parity doublet model, there are two linear couplings of the scalar fields $g^{(1)}_\sigma, g^{(1)}_\zeta$ to the baryonic fields, where the second one generates the mass splitting of the parity doublet states.
%The scalar meson interaction, driving the spontaneous breaking of the chiral symmetry,
%can be written in terms of SU(3) invariants
%$I_2 = (\sigma^2+\zeta^2) ,~ I_4 = -(\sigma^4/2+\zeta^4)$ and $I_6 = (\sigma^6 + 4\zeta^6)$ as:
%
%\begin{equation}
%V = V_0 + \frac{1}{2} k_0 I_2 - k_1 I_2^2 - k_2 I_4 + k_6 I_6 ~,
%\label{veff}
%\end{equation}
%
%where $V_0$ is fixed by demanding a vanishing potential in the vacuum. 
In order to avoid introducing too many parameters, we assume that the splitting of the various baryon species and their respective parity partners has the same value for all baryons. The hyperonic vector interactions are tuned to generate phenomenologically acceptable optical potentials of the hyperons in ground-state nuclear matter. The relevant parameters have been tabulated in Ref. \cite{Mukherjee:2016nhb}. For the calculations done in this paper, the parametrization is kept similar to that used earlier, the only difference being the omission of the baryon decuplet, and other higher resonances, from the particle mixture, which simplifies the discussion without major quantitative changes.

%This is a logical approximation because, with an increase in the magnitude of the strangeness-chemical potential, the system becomes increasingly stacked with hyperons and strange-quarks. The higher resonances do not couple directly to the fields, and hence, they do not have any direct contribution to driving the transitions. They do have an indirect effect on the transitions, though, when present in considerable numbers, through the excluded-volume corrections \cite{Cleymans:1992jz}. But in an environment filled with heavy hyperons and strange-quarks, even the full baryon octet remains unfilled. Thus, to simplify calculations; without sacrificing too much generality; the higher resonances are excluded, as they can be considered non-existent for all practical purposes, under the current circumstances.

In many-body systems, like those resulting from HICs, a chemical potential can be associated with each of the conserved charges of the system. In an HIC's case, the corresponding charges are the baryon-number, isospin and strangeness \cite{bass2000clocking,karsch2011probing,bazavov2012freeze}, because of the short time elapsed between the formation of the fireball and the chemical, and kinetic, freeze-outs, assuming strangeness equilibration. During this time, only strong interactions play an important role, while electroweak interactions are practically negligible. As first argued in Ref. \cite{Rafelski:1982pu}, strangeness might be abundantly produced in the deconfined phase through gluon-gluon fusion, during the early stages of the system's evolution. The strange quarks are later rapidly redistributed in the hadronic phase, via multi-mesonic interactions, when the system is close to the transition \cite{Greiner:2000tu}.

Although the total strangeness of the entire system (fireball) remains zero throughout its formation and evolution, local distributions of non-zero strangeness (and anti-strangeness) regions could be formed as a result of  fluctuations, resulting in a non-uniform distribution of strangeness within the system \cite{Torrieri:2006ag,greiner1988creation,schaffner1997properties,schaffner1997detectability}. These local sub-systems can be considered as being in thermal equilibrium with the rest of the system; since they are considerably smaller in size compared to the entire system. Thus, they can be adequately described by a grand-canonical ensemble.

The pressure ($P$) for such a thermalised system can be written as:

\begin{equation}
P = - E + TH + B_j \mu_{B_j} + S_j \mu_{S_j} + I_j \mu_{I_j} ~,
\label{grand_can_pot}
\end{equation}

with $E$, $T$, $H$, $\mu$, $B_j$, $S_j$ and $I_j$ representing the energy, temperature, entropy, chemical potential, baryon-number, strangeness and isospin, respectively, of the different particle species; and the relative sign between $B$ and $S$ being always negative. In the quark phase, strange-quarks (or anti-quarks) carry a baryon number of $1/3$ (or $ - 1/3$).

For the purpose of this paper isospin effects are not considered and  Eqn. (\ref{grand_can_pot}) reduces to:

\begin{equation}
P = - E + TH + B_j \mu_{B_j} + S_j \mu_{S_j} ~.
\label{modmui0}
\end{equation}

\begin{figure}
\centering
\includegraphics[width=0.35\textwidth,angle=270]{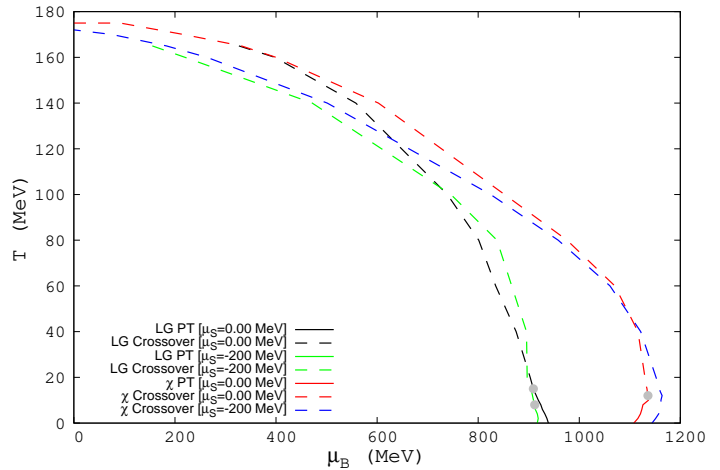}
\caption{$T-\mu_B$ phase-diagram, showing the LG and chiral transitions at $\mu_S =$ 0 and $-200$ MeV.}
\label{muscrit}
\end{figure}

Due to a non-zero $\mu_S$, depending on the sign, hyperon thresholds are lowered to values below, or close to, the masses of the non-strange baryons. Thus, the hyperonic particles appear in the system at smaller values of $\mu_B$, as compared to the case of 
$\mu_S = 0$. The hyperons produced have two non-zero quantum numbers ($B_j$ and $S_j$) and chemical potentials 
($\mu_{B_j}$ and $\mu_{S_j}$). These changes naturally drive the first-order, nuclear Liquid-Gas (LG) transition to lower values of $\mu_B$, as shown in Fig. \ref{muscrit}. At $\mu_S = -200$ MeV, not only is the 
LG transition line shifted to the left, but also its critical end-point ($T_{\rm CEP}$,$\mu_{B_{\rm CEP}}$) is lowered along the 
$T-$axis; from 15 MeV (for $\mu_S = 0$ MeV) to 8 MeV (for $\mu_S = -200$ MeV), weakening the phase boundary to a crossover, earlier than that with a vanishing $\mu_S$. 
%The critical temperature, $T_C$, at a vanishing $\mu_B$, for $\mu_S = -200$ MeV, is observed to be 172 MeV. This is in good agreement with LQCD results; which predict a $T_C$ of $175 \pm 6$ MeV, at $\mu_B = 0$ MeV \cite{AliKhan:2000wou,Bernard:2004je,Fodor:2004nz,Fodor:2001pe,Karsch:2000kv,Aoki:2006we,Aoki:2009sc}. 
It is also evident from the figure that the chiral  first-order transition weakens with larger negative values of $\mu_S$ and disappears completely below $\mu_S = -200$ MeV, giving way to a smooth crossover transition, for the full range of temperatures. 

\begin{figure}
\centering
\includegraphics[width=0.35\textwidth,angle=270]{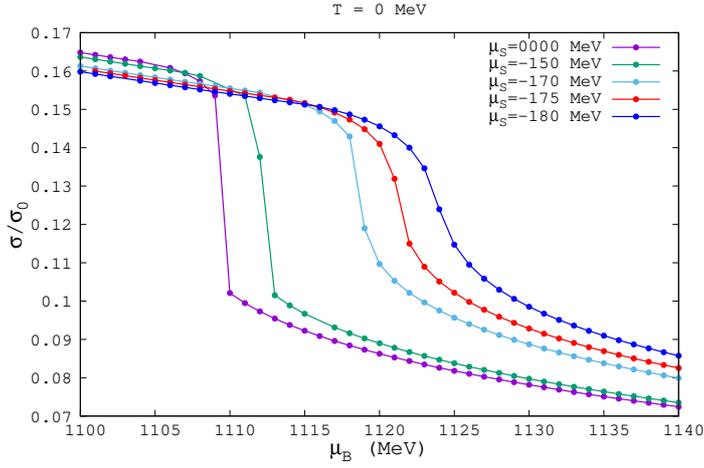}
\caption{Normalised chiral condensate, as a function of $\mu_B$, for different values of $\mu_S$, at $T=0$.}
\label{sigma_mub}
\end{figure}

\begin{figure}
\centering
\includegraphics[width=0.35\textwidth,angle=270]{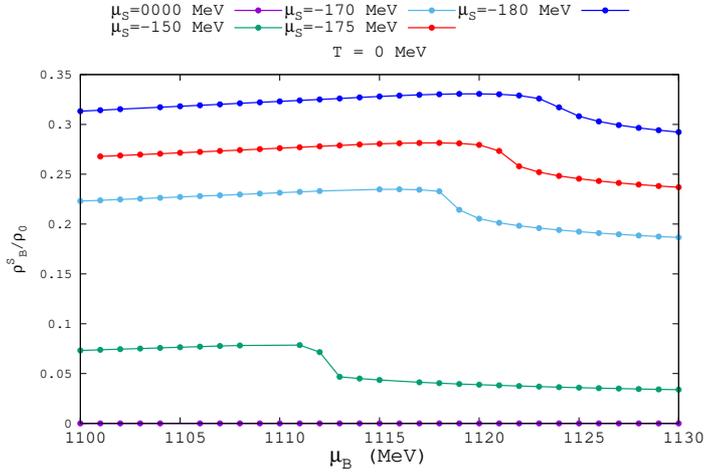}
\caption{Relative abundance of hyperons, as a function of $\mu_B$, for different values of $\mu_S$, at $T=0$.}
\label{brho}
\end{figure}

\begin{figure}
\centering
\includegraphics[width=0.35\textwidth,angle=270]{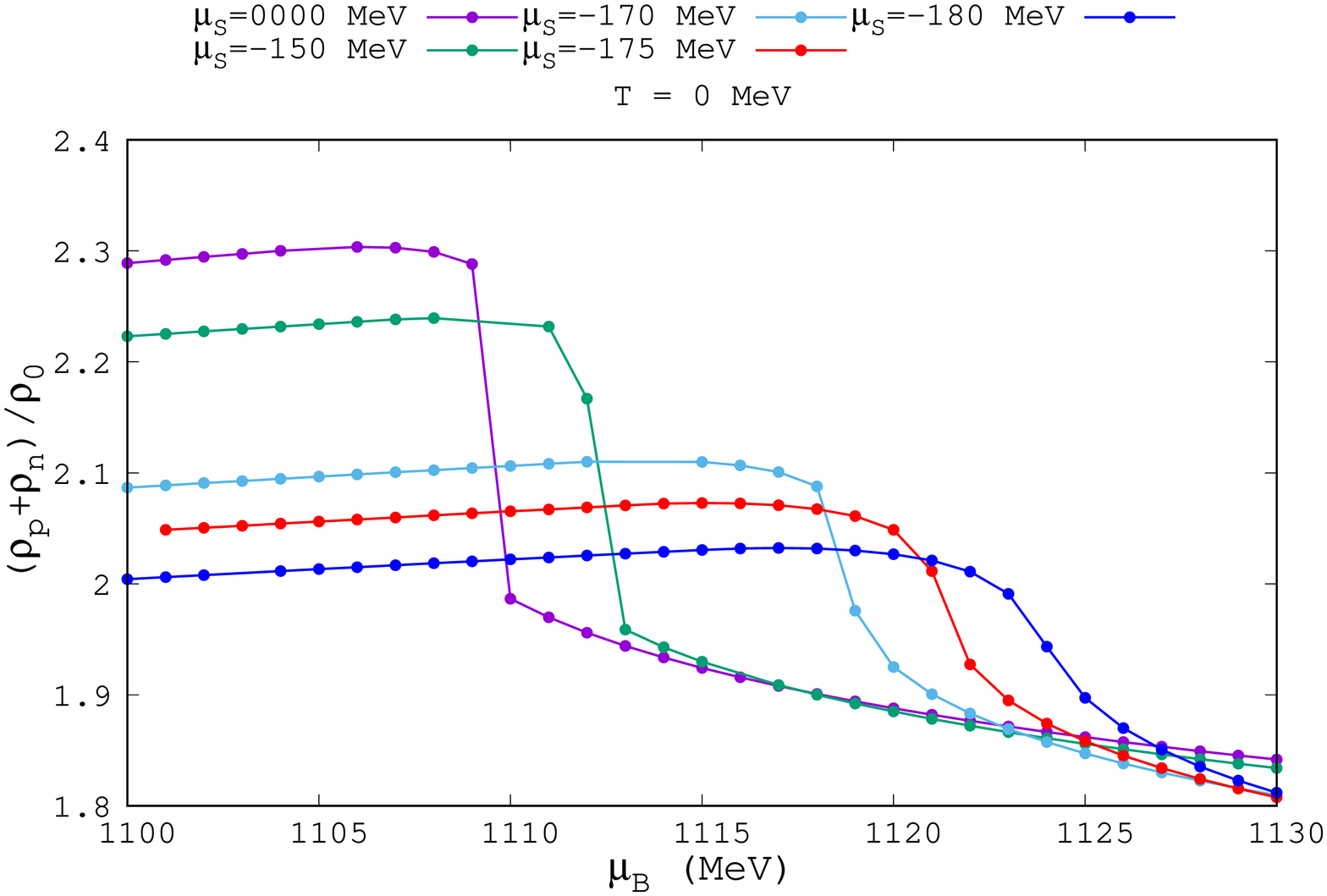}
\caption{Relative abundance of non-strange baryons, as a function of $\mu_B$, for different values of $\mu_S$, at $T=0$.}
\label{pnrho}
\end{figure}

\begin{figure}
\centering
\includegraphics[width=0.35\textwidth,angle=270]{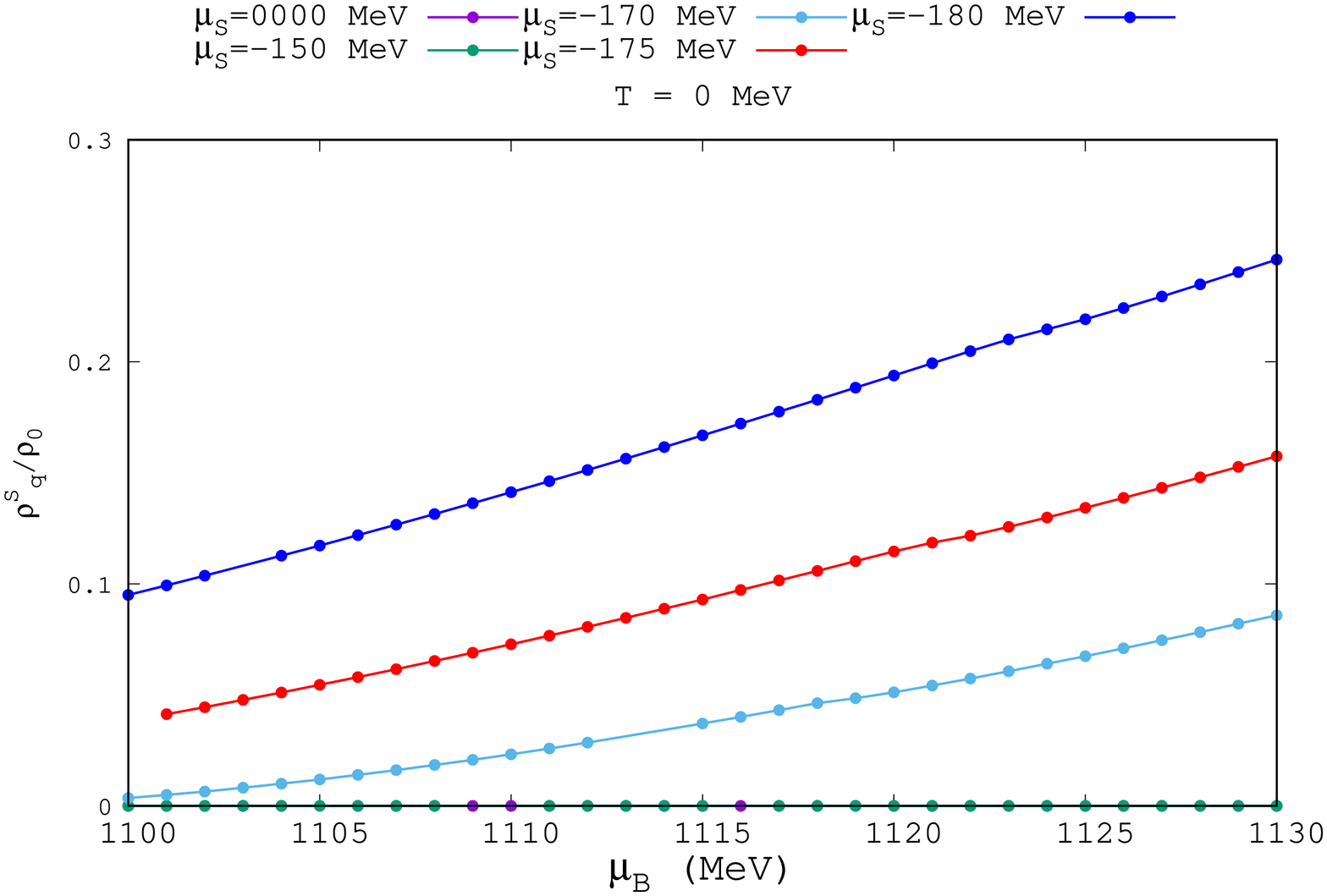}
\caption{Relative abundance of strange-quarks, as a function of $\mu_B$, for different values of $\mu_S$, at $T=0$.}
\label{qrho}
\end{figure}

\begin{figure}
\centering
\includegraphics[width=0.35\textwidth,angle=270]{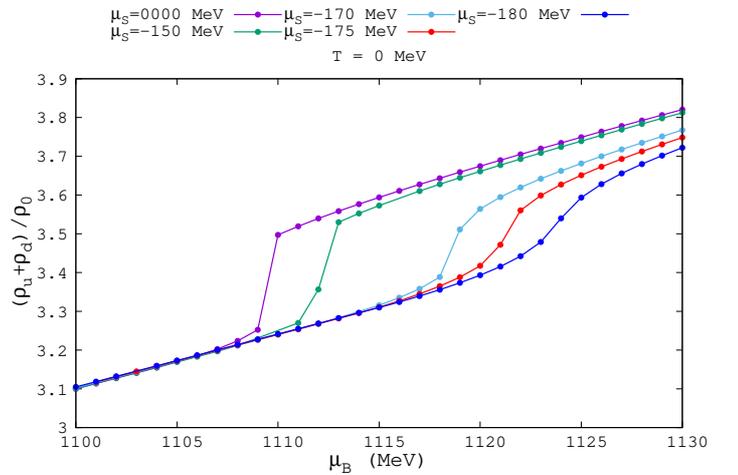}
\caption{Relative abundance of non-strange quarks, as a function of $\mu_B$, for different values of $\mu_S$, at $T=0$.}
\label{udrho}
\end{figure}

In Fig. \ref{sigma_mub} the normalised scalar field ($\sigma/\sigma_0$) is plotted as a function of $\mu_B$ at $T=0$ MeV. One can observe that the chiral first-order transition actually vanishes at $\mu_S = -175$ MeV. As can be seen in Figs. \ref{brho}, \ref{pnrho}, \ref{qrho} and \ref{udrho}, the chiral condensate is intimately related to the net-baryon density, and hence, the change in either variable can be used to define the transition~\cite{Mukherjee:2016nhb,Walecka:1974qa,
Bender:2003jk,Cohen:1991nk}. From the figures, one arrives at the immediate conclusion that, with increasing $\left| \mu_S \right |$, these quantities exhibit progressively shallower jumps near the transition, pointing to a weakening of the first-order phase transition. The increase of higher-mass hyperons in the hadronic phase reduces  the relative abundance of the lower-mass, non-strange baryons (Figs. \ref{brho} and \ref{pnrho}). 
Fig. \ref{qrho} shows that the strange-quark degrees-of-freedom, already present in the system before the transition (in the hadronic phase), increase, with an increase in $|\mu_S|$; causing the relative contribution of the lighter, non-strange quark degrees-of-freedom to decrease (Fig. \ref{udrho}). By significant couplings to the much stiffer strange-quark condensate $\zeta$, the hyperons gradually push the chiral transition to higher values of $\mu_B$. Since the transition is signalled by an abrupt decrease in $\sigma$, to which the nucleons couple more strongly, a lower concentration of these non-strange baryons at moderate $\mu_B$ causes the hadronic phase to survive much longer than that for a vanishing $\mu_S$. 
Moreover, this suppression of the non-strange baryons causes a smoothing of the transition, even at lower values of $|\mu_S|$, as seen in Fig. \ref{sigma_mub}. When the concentration of strange-quarks in the hadronic phase increases further, with higher values of $|\mu_S|$ (Fig. \ref{qrho}), the degrees-of-freedom do not change as drastically across the chiral transition, resulting in a smooth crossover, instead of a sharp first-order, for all strangeness-chemical potentials $ \leq -175$ MeV and temperatures $ \geq 0$ MeV.

%\textcolor{blue}{AM: Why does the LG transition shift to the left, while the chiral shifts to the right?}

\begin{figure}
\centering
\includegraphics[width=0.35\textwidth,angle=270]{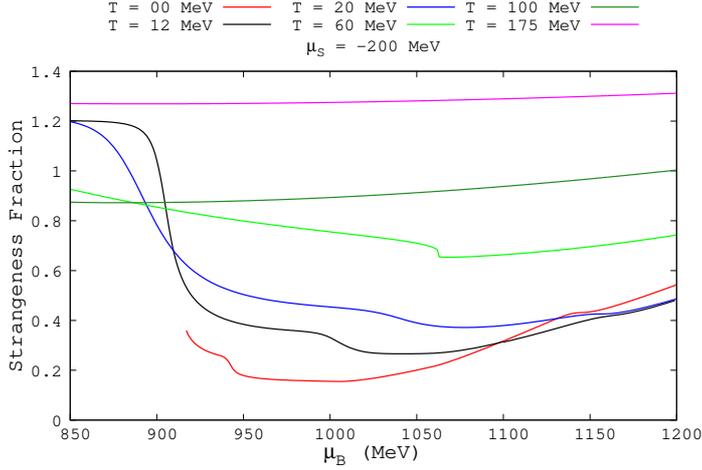}
\caption{Strangeness fraction, as a function of $\mu_B$, for different values of $T$, at $\mu_S = - 200$ MeV.}
\label{fs_200musm}
\end{figure}

\begin{figure}
\centering
\includegraphics[width=0.35\textwidth,angle=270]{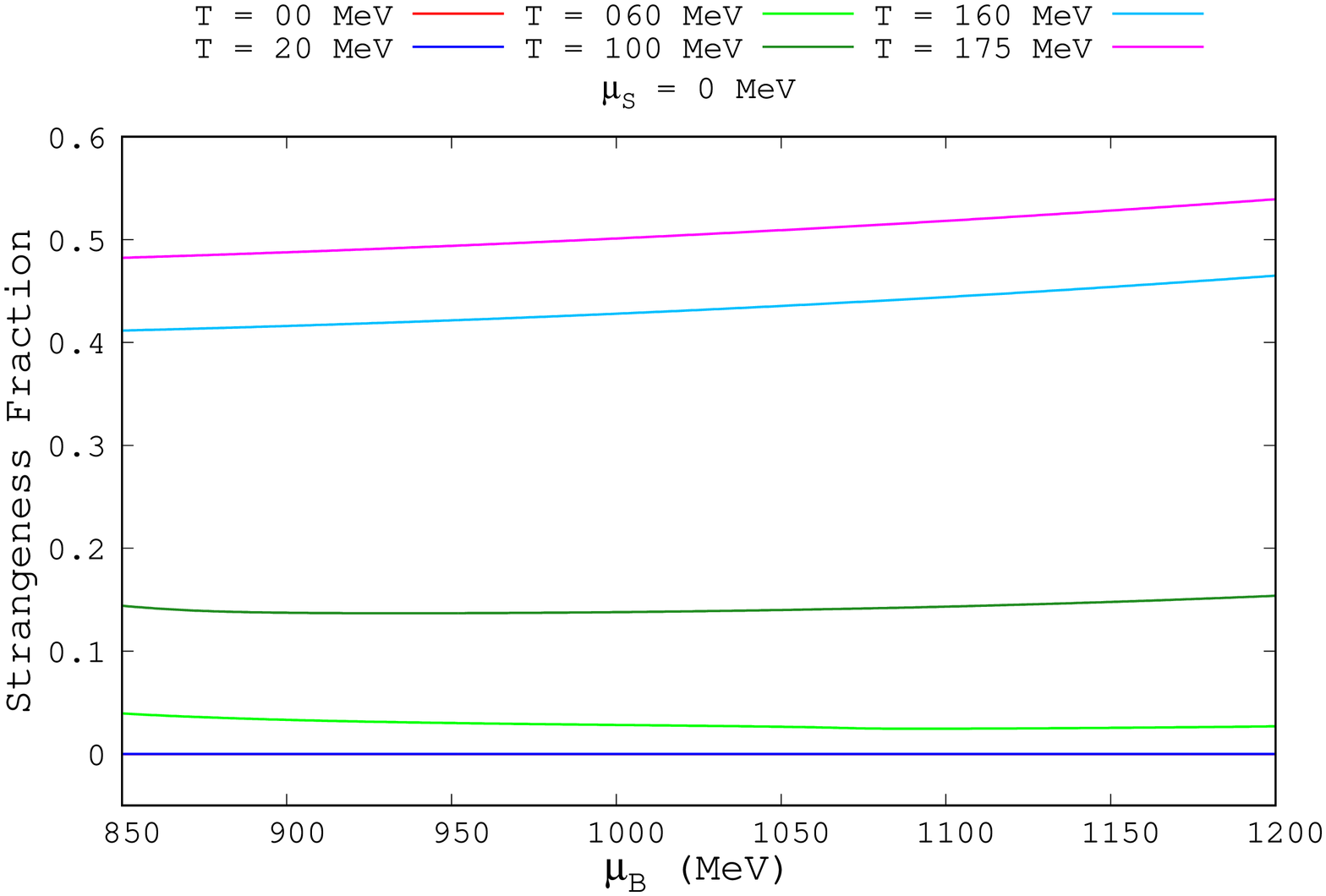}
\caption{Same as Fig. \ref{fs_200musm}, at $\mu_S = 0$ MeV.}
\label{fs_0musm}
\end{figure}

\begin{figure}
\centering
\includegraphics[width=0.35\textwidth,angle=270]{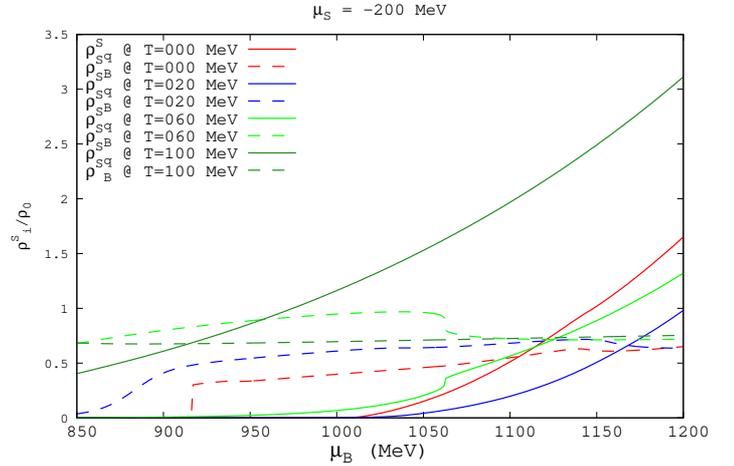}
\caption{Relative abundances of strange-quarks and hyperons, as functions of $\mu_B$, for different values of $T$, at $\mu_S = -200$ MeV.}
\label{part_200musm}
\end{figure}

\begin{figure}
\centering
\includegraphics[width=0.35\textwidth,angle=270]{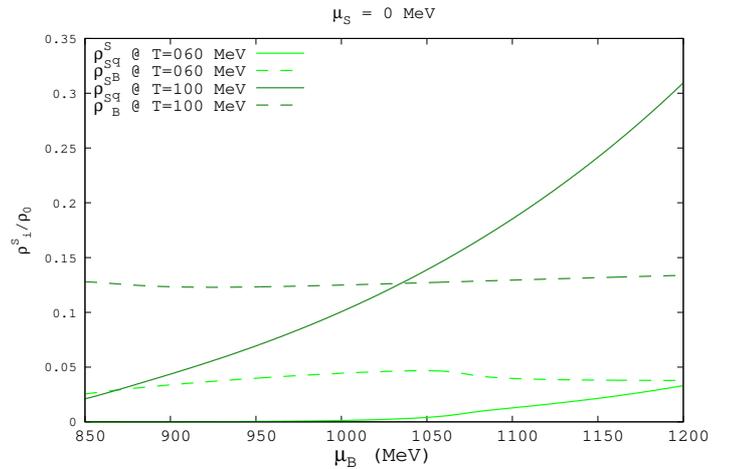}
\caption{Same as Fig. \ref{part_200musm}, at $\mu_S = 0$ MeV. The curves corresponding to temperatures less than 60 MeV are negligibly close to zero and have not been shown.}
\label{part_0musm}
\end{figure}

\begin{figure}[t!]
\centering
\includegraphics[width=0.35\textwidth,angle=270]{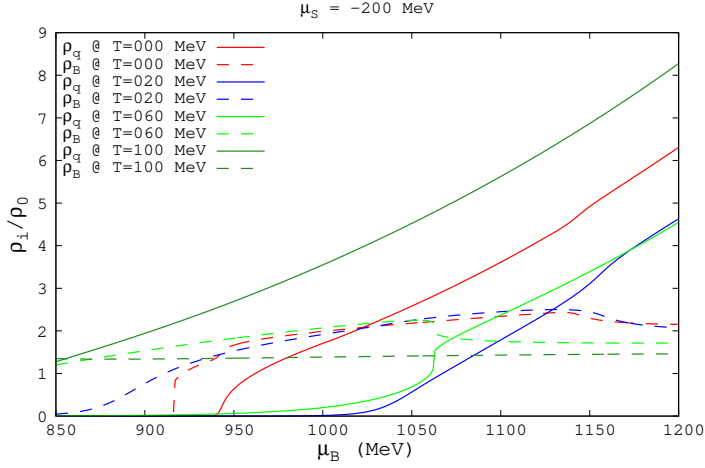}
\caption{Relative abundances of quarks and baryons, as functions of $\mu_B$, for different values of $T$, at $\mu_S = -200$ MeV.}
\label{200relab}
\end{figure}

\begin{figure}
\centering
\includegraphics[width=0.35\textwidth,angle=270]{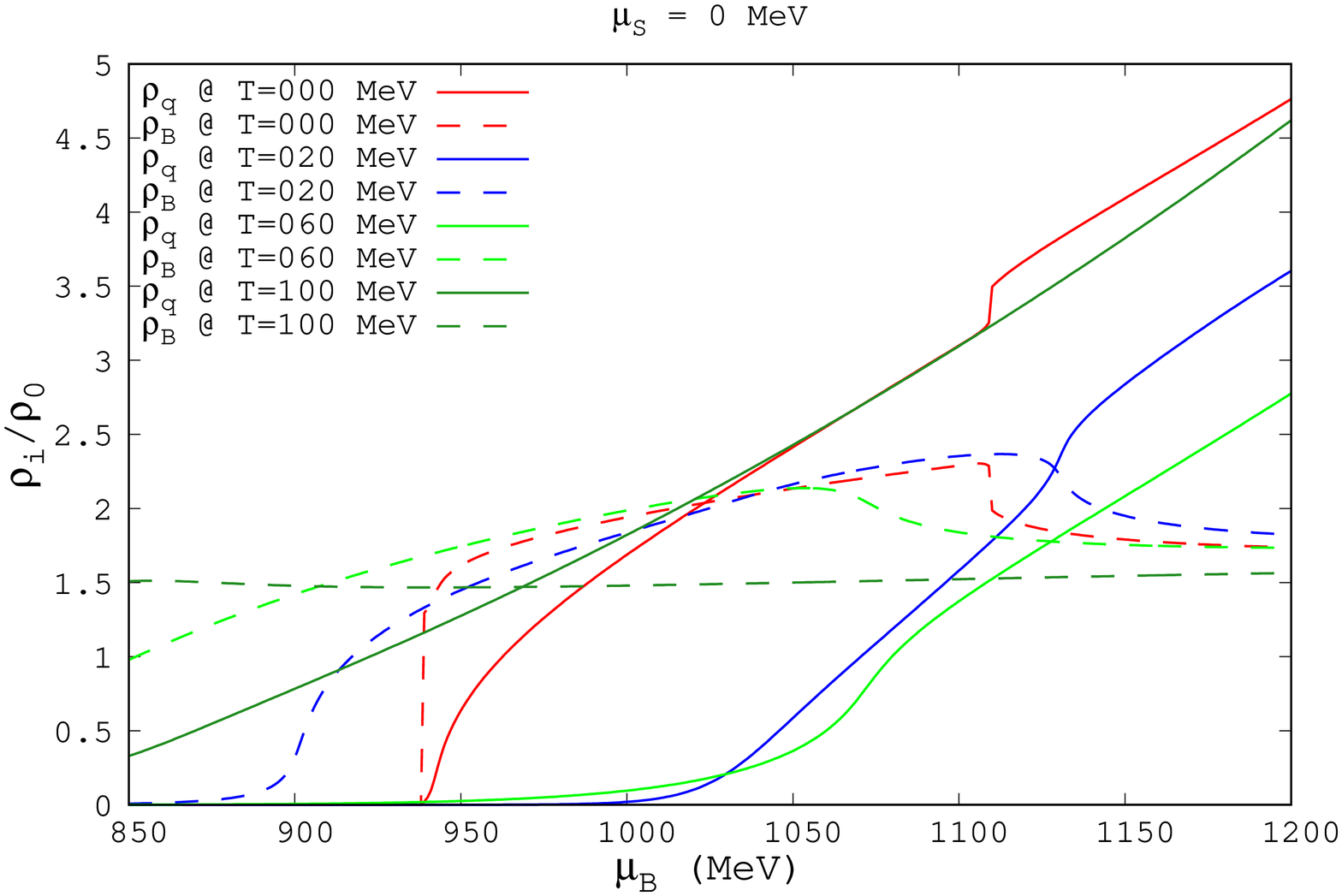}
\caption{Same as Fig. \ref{part_200musm}, at $\mu_S = 0$ MeV.}
\label{0relab}
\end{figure}

In Figs. \ref{fs_200musm} and \ref{fs_0musm}, the strangeness fraction ($f_S$); defined as:

\begin{equation}
f_S = \frac{\rho_B^S}{\rho_B}~;
\end{equation}

is plotted against $\mu_B$, at different temperatures, for $\mu_S = -200$ MeV and 0 MeV, respectively. The baryon number density $\rho_B^S$ includes contributions from both quarks and baryons. In Figs. \ref{part_200musm} and \ref{part_0musm}, the relative abundances of the strange-quarks and hyperons are plotted, while in Figs. \ref{200relab} and \ref{0relab}, the normalised particle-number-densities; for all quarks and baryons, at different temperatures; are plotted against $\mu_B$, for constant values of $\mu_S$ ($-200$ MeV and 0 MeV, respectively).

\begin{figure}[!t]
\centering
\includegraphics[width=0.34\textwidth,angle=270]{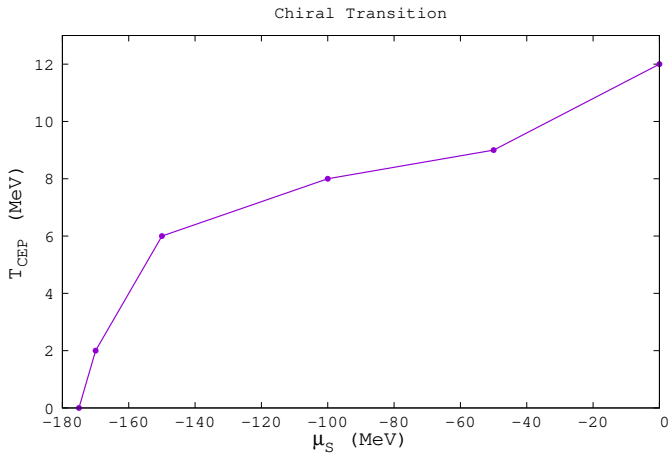}
\caption{Critical end-point temperature, for the chiral transition, as a function of $\mu_S$.}
\label{tcep}
\end{figure}

As is amply evident from Fig. \ref{fs_0musm}, for $\mu_S = 0$, the system lacks the rich structure, at lower temperatures, visible in Fig. \ref{fs_200musm}. Moreover, the curve for $T = 0$ MeV in this figure is buried beneath the $T = 20$ MeV curve. So there is no evidence of any structure between the temperatures 0 and 20 MeV; as corroborated by Fig. \ref{part_0musm}.

In Fig. \ref{fs_200musm}, the $T = 0$ MeV curve begins exactly after the first-order LG transition, at $\mu_B \sim 920$ MeV. This sudden appearance of strangeness can be attributed to the introduction of the $\Lambda$ and $\Xi$ hyperons to the system - along with other baryons - as can be seen in Fig. \ref{part_200musm}, thereby making both $\rho_B^S$ and $\rho_B$ non-zero. The shoulder-like dip at $\mu_B \sim 940$ MeV is the result of the early onset of the up- and down-quarks, as seen in Fig. \ref{200relab}.

For the $T = 12 \textrm{ and } 20$ MeV curves in Fig. \ref{fs_200musm}, $f_S$ decreases drastically after the LG transition. This is because, right after the transition, there is a sudden rise in $\rho_B$, while the strange-particle contribution $\rho_B^S$ does not rise as much, due to the higher masses of the hyperons, which change relatively less strongly across the transition. This drives down the fraction of strangeness in the system, which is slowly revived as the hyperons start increasing in abundance with increasing $\mu_B$, as is evident from the gradual rise of $\rho_B^S$, for $T = 12 \textrm{ and } 20$ MeV, in Fig. \ref{part_200musm}. With the appearance of the up- and down-quarks (Fig. \ref{200relab}), at around $\mu_B \sim 1000$ MeV, $f_S$ again experiences a slight dip in value. The third and final dips, observed at $\mu_B \sim 1140$ MeV, are caused by the chiral first-order transition (Fig. \ref{muscrit}), which is not as sharp compared to the nuclear LG transition. As seen in Figs. \ref{part_200musm} and \ref{200relab}, the quarks start dominating the composition of the system, as $\mu_B$ increases, from this point onwards.

The kink in the $T = 60$ MeV curve (Fig. \ref{fs_200musm}) is 
caused by the chiral crossover transition, as evident from Figs. \ref{muscrit}, \ref{part_200musm} and \ref{200relab}. Expectedly, after the transition into the quark sector, the relative abundance of baryons decreases w.r.t. quarks; only in this case, the decrement is much smoother, and smaller, as compared to a first-order transition.

For $T = 100 \textrm{ and } 175$ MeV, the respective chiral crossover transitions occur at $\mu_B \sim 840 \textrm{ and } 0$ MeV (cf. Fig. \ref{muscrit}). As expected, the corresponding $f_S$ curves in Fig. \ref{fs_200musm} are monotonously increasing functions of $\mu_B$, for the range of values (850$-$1200 MeV) considered. 

The figures, in addition to showing the disappearance of the chiral first-order transition at higher $\mu_S$ values, showcase the effect that $\mu_S$ has on the system as a whole. The fraction of strangeness in the system, driven by the growing relative abundance of the hyperons and strange-quarks, increases rapidly with $\mu_B$ in Figs.  \ref{fs_200musm}, \ref{part_200musm} and \ref{200relab}. They also grow to much higher values, as compared to what they attained with a zero strangeness-chemical potential, for similar values of $\mu_B$ (Figs. \ref{part_0musm} and \ref{0relab}). A non-zero $\mu_S$ also results in an early onset of the aforementioned strange-particles, as evidenced by the shifting of the kink; corresponding to the chiral transition; in Fig. \ref{fs_200musm}, to progressively lower values of $\mu_B$, with an increase in temperature.

In the case of $\mu_S = 0$ MeV, the strangeness-fraction is observed to be either monotonously increasing, or remaining fairly constant, with $\mu_B$; for all temperatures in Fig. \ref{fs_0musm}. This is to be expected, however, since from Figs. \ref{part_0musm} and \ref{0relab}, it is clear that the transitions are primarily driven by the changes in the relative abundances of the non-strange quarks and baryons. But, even in this case, with an increase in temperature, strange-particles with baryon numbers do start to come in due to strange-mesons, in particular kaons.
% as the aforementioned sub-systems of non-zero strangeness, and anti-strangeness, are formed in a high-density and moderate-temperature environment. 
 This explains the existence of a non-zero $f_S$, which increases with an increase in temperature of the system, for a zero strangeness-chemical potential. The slight dip in $f_S$, at $T = 60$ MeV, is again caused by a sudden increase in $\rho_B$ across the chiral transition (Fig. \ref{muscrit}), with $\rho_B^S$ not being able to change as rapidly.

%\begin{figure}
%\centering
%\includegraphics[width=0.335\textwidth,angle=270]{mucep_mus.eps}
%\caption{Critical end-point baryo-chemical potential, for the chiral transition, as a function of $\mu_S$.}
%\label{mucep}
%\end{figure}

%\textcolor{red}{AM: The content of the following paragraph is incomplete. So, PLEASE DO NOT READ!}

In Fig. \ref{tcep}, the critical end-point temperature is plotted as a function of $\mu_S$. As expected, a considerable, gradual decrease in $T_{\rm CEP}$ is observed, with an increase in the magnitude of $\mu_S$. This re-emphasises the fact that the strangeness-chemical potential directly affects the LG and chiral transitions. The hadronic phase is dominated by hyperons, as $\mu_S$ increases in magnitude, suppressing other baryons and resulting in an early onset of both light ($u-$ and $d-$) and strange quarks; which go on to become a quark state at hight densities.

%\textcolor{blue}{$\rightarrow$ include conclusion, further discussions and outlook here}
In conclusion, in this model investigation of quark-hadron systems, one could see that the QCD phase diagram can be significantly affected by a non-zero strangeness-chemical potential, 
%As explained earlier, it is not outlandish to expect such a situation in an HIC, where the temperature and density provide conditions ripe for the formation of small, non-zero strangeness (and anti-strangeness) sub-systems. Due to the short time-scales associated with the evolution of the system, the hyperons formed in these sub-systems are not expected to decay into non-strange hadrons~\cite{schaffner1993strange,schulze1998hyperonic,nakamura2010review,yao2006review,beringer2012pdg} and can, therefore, affect the particle output detected from the collision. Tracing the particle-number distributions, and their fluctuations, back to the initial stages of the system, the conjectured influence of the strangeness, on the evolution of the system, can be confirmed.
changing the chiral transition from a first-order to a smooth crossover. The critical endpoint in this model appears at low temperatures, which makes such an effect difficult to be directly observed in heavy-ion collisions, but it could have an impact in the higher-temperature smooth transition region as well. Another strangeness-enriched situation is the beta-equilibrated matter in a neutron star, which was investigated in Ref. \cite{Mukherjee:2017jzi}.

A related study could include explicit isospin effects on the behaviour of high-density and high-temperature QCD systems and the role that isospin plays in the system evolution. The fact that non-uniform distributions of isospin in the system are usually very small, is partly compensated by the possibility of experimentally measurable observables. These non-zero isospin-chemical potential effects, and their consequences as experienced by the Q$\chi$P model, will be investigated in upcoming projects. 

The authors would like to thank the Board of Research in Nuclear Sciences (BRNS), India; the Helmholtz International Centre (HIC) for FAIR and the Bundesministerium f{\"u}r Bildung und Forschung (BMBF), Germany; for their support. They would also like to express their gratitude towards Dr. J. Steinheimer, for his help with the discussions. The calculations for this paper were carried out at the Center for Scientific Computing (CSC)  in Frankfurt, Germany.

%\section{Conclusions \& Outlook}
%\label{sec:concout}

%-------------------------------------------------------------------------------------------
%BIBLIOGRAPHY
%-------------------------------------------------------------------------------------------

%\bibstyle{apsrev4-1}
\bibliographystyle{elsarticle-num}
\bibliography{bibnew}

%-------------------------------------------------------------------------------------------

\end{document}